\documentstyle[aps,multicol,eqsecnum,epsf]{revtex}

\begin{document}
\draft

\title{Spontaneous Stratification in  Granular Mixtures}

\author{Hern\'an A. Makse,$^1$ Shlomo Havlin,$^{1,2}$ Peter R. King,$^3$
and 
H. Eugene Stanley$^1$}

\address{$^1$Center for Polymer Studies and Physics Dept., Boston
University, Boston, MA 02215 USA\\ 
$^2$Minerva Center and 
Department of Physics, Bar-Ilan University, Ramat Gan, Israel\\
$^3$BP Exploration Operating Company Ltd.,
Sunbury-on-Thames, Middx., TW16 7LN, UK}
\date{ Nature {\bf 386}, 379-381 (1997)}
\maketitle
\begin{abstract}
Granular materials \cite{bagnold2,nagel,bideaux,mehta,wolf}
size segregate
when  exposed to external
periodic perturbations such as vibrations
\cite{williams,rosato,herrmann,knight,zik,duran,warr}.
Moreover,
mixtures of grains of different sizes spontaneously   
segregate  in the absence of external
perturbations: when 
a  mixture is simply poured onto a pile, the large grains
are 
more likely to be found near
the base, while the small grains are more likely to be near the top
\cite{brown,bagnold0,drahun,fayed,savage1,savage2,savage3,meakin}.  
Here, we report a novel spontaneous phenomenon
arising 
when we pour a mixture  between two vertical plates: the
mixture spontaneously stratifies into alternating layers of small and large
grains whenever 
the large grains have larger angle of
repose than the small grains. 
In contrast, we find  only spontaneous segregation when 
the large grains have smaller angle of repose  than the small grains.
The stratification is related to the occurrence of
avalanches; during each avalanche the grains comprising the avalanche
spontaneously stratify into a pair of layers, 
with the small grains forming a sublayer underneath the layer of
large grains.
\end{abstract}
\vfill
\begin{multicols}{2}
Our experimental system consists of a vertical ``quasi-two-dimensional''
 cell  with a  gap of 5 mm separating two
transparent plates (made of plexiglass, or of glass) measuring 300 mm
$\times$ 200 mm (see Fig. \ref{experiment}{\it a}). We choose this
quasi-two-dimensional geometry since, by using this setup, the internal
features of the avalanche process can be easily visualized, both
statically and dynamically. To avoid the effects of electrostatic
interaction with the wall, the wall is cleaned with an antistatic
cleaner.

In a first series of experiments, we close the left edge of the cell
leaving the right edge free, and we pour, near the left edge, an
equal-volume mixture of white glass beads  (mean size 0.27 mm, 
spherical shape, and repose angle $26^o$), and red sugar crystals (typical
size 0.8 mm, cubic shape, and repose angle $39^o$).   ~Figure
\ref{experiment}{\it a} shows the result of the first series of
experiments. We note two features:

\begin{itemize}
\item[(i)] {\it Spontaneous Stratification}. We see the formation of
alternating layers consisting of small and large grains---with a
``wavelength'' of about 1.2 cm.

\item[(ii)] {\it Spontaneous Segregation}. We find that the smaller grains
segregate near the left edge and the larger grains segregate furthest
from it and near the base
\cite{brown,bagnold0,drahun,fayed,savage1,savage2,savage3,meakin}.  
\end{itemize}

In a second series of experiments, we confirmed the results of these
initial experiments by testing for  stratification and
segregation using a mixture of grains of same density, consisting
of fine sand (typical size 0.4 mm) and coarse sand (typical size 1 mm), 
suggesting that the density of the grains may not play an important role
in stratification.  

In all the above experiments we used 
mixtures composed of two types of grain
with different shape, and therefore with different
angles of repose. In particular we obtain stratification
(plus segregation) when we use larger cubic grains and smaller spherical
grains:
the angle of repose of
the large species is then larger than the angle of repose  of the small
species.  
Otherwise we obtain only segregation and not stratification when the
large grains are less faceted than the small grains, i.e., the large
grains
have smaller angle of repose than the small grains.

To confirm this, we  performed a series of experiments using  mixtures of
irregular shaped 
sand grains (repose angle $35^o$, and mean size 0.3 mm), 
and spherical glass beads (repose angle $26^o$ {\it smaller}
than the repose angle of the sand grains). 
We find that stratification (plus
segregation) occurs for two different experiments
using spherical beads of 
size 0.07 mm and 0.11 mm (so that the larger grains have larger
repose angle).
In contrast, we obtain only segregation {\it but not
stratification} for two experiments using spherical  beads of
size 0.55 mm and 0.77 mm (so that the larger grains have smaller repose angle).
 In all  cases the segregation
of grains occurs with the smaller grains being found near the left
edge of the cell and the larger grains near the base of the cell.
These results suggest that the phenomenon of  segregation is always
expected when pouring a granular mixture of grains of different sizes,
no matter the values of the angles of repose of the species.  However,
the phenomenon of  stratification is only expected when the large
species have larger angle of repose than the small species.

Additionally, we performed a series of experiments in which we
find similar stratification  by using different
mixtures of differing size ratio between large and small grains (1.66,
2.1, 2.25, 3.25, and 6.66),
suggesting that the  phenomenon occurs for a broad
regime of grain size ratios.
We find a similar stratification when we
double the gap between the vertical plates of the cell and simultaneously
double the flow rate
of grains.

We propose a physical mechanism responsible for the observed
stratification that is related to the fact that not one but rather
a pair of layers is formed in the course of each avalanche. 
When the flow  of grains reaches the base of the pile, we find that the
grains develop a profile characterized by a well-defined ``kink'', at
which the grains are stopped (see Fig. \ref{experiment}{\it b}); 
but we find that the small grains stop first, so a pair of
layers forms with the 
small grains {\it underneath} the large
grains.
As more
grains are added, 
the  kink appears to move upward in the direction
opposite to the flow of grains. Once the kink reaches the top,
the pair
of layers is complete and the cycle is then repeated:
a new avalanche occurs,  the kink 
develops, and a new pair of layers forms.

The ``wavelength'' 
of a pair of layers $\lambda$ can be determined by the mean
value of the downward velocity $v$ of the rolling grains during an
avalanche, the upward velocity $v'$ of the kink, and the thickness of
the layer of rolling grains $R_0$ during the avalanche.   If the
volume of grains in an avalanche scales approximately as the volume of
grains in a well-formed kink, we predict $\lambda \simeq R_0 (v+v') /
v'$, 
and we
confirm this relation experimentally.

To test this physical mechanism by computer simulation we consider a
mixture comprising small grains of width one pixel and of height $H_1$,
and large grains, also of width one pixel but of height $H_2 > H_1$.  To
generate an equal-volume mixture, we randomly drop a small grain with
probability $p\equiv H_2/(H_1+H_2)$, and drop a large grain with
probability $1-p$ (see Fig. \ref{dynamics}).
 
In critical phenomena, it is often useful to first develop a ``mean
field'' type model (in which, for example,  spin 
orientation is determined by a
macroscopic variable, the net magnetization), before devising models in
which spin orientation is determined by the microscopic quantities such
as the orientations of the other spins comprising the system. In this
spirit, we focus first not on the ``microscopic'' grain motions, but
rather on the ``macroscopic'' angle of the sandpile, whose value
alternates in time between the maximum angle of stability $\theta_m$
which defines the onset of an avalanche, and the angle of repose
$\theta_r$ which defines the end of the avalanche
\cite{bagnold3,nagel2}. Using this model (described in the legend of
Fig. \ref{dynamics}), 
we find a morphology that displays both segregation and
stratification.

In addition to the simplest ``mean field'' approach, we develop a model
\cite{makse} 
in which
we treat the individual grain motion in accord with microscopic rules
that depend not on the macroscopic angle of the sandpile, but rather on
the local angles formed between each grain and its neighbors.
Specifically, the dynamics of the small and large rolling grains
are governed
by the critical angles of repose 
corresponding
to the interactions between the rolling grain
and the static grains of the sandpile surface. 
This model incorporates the experimental fact that grains
segregate because large grains roll down more easily on top of
small grains than small grains on top of large grains (for rolling {\it
large} grains on top of a surface of small grains
the surface appears smoother than for rolling {\it small}
grains rolling on top of a surface of large grains).
Thus the model
correctly predicts that the small grains form a sublayer beneath the
large grains. 
We find
stratification, as in the simplest ``mean field'' model, and also find that the
profile of the sandpile displays a kink at which rolling grains are
stopped---just as in the experiment.

Next we test the above principles by generalizing from two
grain sizes to three. 
The  experiment results in
stratification with three layers, with the finest grains on the
bottommost of each triplet of layers and the coarsest grains on the
topmost layer
(see Fig. \ref{experiment}{\it c} where the same
experimental setup of Fig. \ref{experiment}{\it a} is used to obtain an
alternation of {\it three} layers of grains of {\it three} different
sizes: 0.15 mm, 0.4 mm, and 0.8 mm). 
Experiments using a
continuum size distribution are ongoing, since geological rock
formations (which also display stratification) generally occur in the
presence of a continuum distribution of grain size.

As another test of the proposed physical mechanism, we note that the case
of spherical grains should not lead to stratification because the angles
of repose of the large and small grains are the same.
We confirm this prediction experimentally.
The case of spherical grains
was also studied by Williams \cite{williams63,williams68,allen};
his results (showing segregation plus a hint of stratification)  
differ from our results (showing only segregation), 
presumably
because his grains were not quite spherical--- i.e.,
 the repose angle of the large and small ``spheres''
were
slightly unequal.

Finally, we note that Boutreux and de Gennes have recently made
 considerable progress \cite{degennes}
in developing a general theoretical framework \cite{bouchaud,pgg}  
to treat the case of
granular flows of two different
grains.
Their conclusions \cite{degennes,bouchaud,pgg}
are consistent with the experiments presented here.

ACKNOWLEDGEMENTS.
We thank T. Boutreux, G. Davies, P.-G. de Gennes, H. J. Herrmann, and S. 
Tomassone for stimulating discussions, and P. Cizeau for collaboration
in stages of this work.

\end{multicols}

\begin{figure}
\narrowtext
\caption{Experimental results: {\bf a,} Typical result of 
the first series of experiments,
 showing the formation of successive layers of fine and coarse grains
(here the white grains are glass beads  of average diameter 0.27 mm, while
the larger grains are sugar crystals of typical size 0.8 mm).
We clean the walls of the cell with antistatic cleaner, in order to avoid
the effects of electrostatic interaction between the grains and the
walls. 
We pour the equal-volume mixture near the left edge between two
transparent vertical plates separated by a gap of 5 mm. 
We obtain stratification with a wavelength $\lambda \simeq 1.2$ 
cm. 
{\bf b,} Close-up photograph of the kink where the grains stop
during an avalanche. The small white grains stop first, and then the 
large red grains; hence the small grains form a sublayer underneath the
large grains.
{\bf c,} 
Stratification obtained using a mixture of three different 
types of grains: nearly spherical glass beads (0.15 mm, angle of repose 
26$^o$), blue sand (0.4 mm, angle of repose 35$^o$),  and red 
sugar crystals (0.8 mm, angle of repose 39$^o$). We notice the grading
(from bottom to top)
in a triplet of layers:
small (white), medium (blue), and large (red) grains.
{\bf d,} 
Close-up of {\bf c,}, field of view 40 mm $\times$ 40 mm. }
\label{experiment}
\end{figure}

\begin{figure}
\narrowtext
\caption{Results of modeling: {\bf a,} 
The dynamics of the simplest ``mean field'' 
model are illustrated by this
example with two sizes $H_1=1$ (white) and $H_2=2$ (red), and threshold
slopes $s_r\equiv \tan\theta_r =2$ and $s_m\equiv\tan\theta_m=3$.
Suppose that, at a given instant, the sandpile is at the
critical slope for repose $s_r$. 
To define the dynamical rules for the arriving grains, we consider the
 slope $s_i \equiv h_{i} - h_{i+1}$, where $h_i$ denotes the
 height of the sandpile at coordinate $i$.
We deposit a grain near the
 first column at the left edge of the lattice, where the actual
column position
is chosen from a narrow Gaussian probability distribution centered at
the wall edge.
The non-zero width of this Gaussian
mirrors the fact that grains
often bounce after reaching the pile.
Newly arriving grains accumulate on the
sandpile profile, following dynamics governed by the critical slope
$s_m$; thus a grain moves from the initial landing point at column $i$ to
column $i+1$ if the slope $h_{i} - h_{i+1}$ is larger than $s_m$,
then moves from column $i+1$ to column $i+2$ if $h_{i+1} - h_{i+2}>s_m$,
and so forth. The grain stops at the first column $k$ with $h_{k} -
h_{k+1} \le s_m$. Another grain is now added, and the same rules are
followed. {\bf b,} This entire process continues until a grain reaches the
substrate at the furthest right column of the pile for first time (grain 8 in
this figure). Now, since $s_i > s_m$ for all columns $i$,
the sandpile has become ``unstable''.
We note that  $s_i$ is calculated considering also the
size
of the rolling grain; as a result, the large grains more readily reach a slope
that exceeds the two critical slopes $s_r$ and $s_m$.
{\bf c, } 
We allow the sandpile
to relax toward the repose slope $s_r$ by moving each of the grains with
slope larger than $s_r$ to the nearest column satisfying $h_{i} -
h_{i+1} \le s_r$. Now the deposition starts again, and we iterate the
algorithm until a large sandpile (of typically $10^5$ grains) is
formed.
 We can obtain stratification with constant ``wavelength''
by stopping the accumulation process 
when a grain reaches for first
time the
column  $l^{1/2}$, where $l$ denotes  the furthest 
right column of the pile.
{\bf d,} Image
obtained with the simplest ``mean field''  model
 (for parameters $H_1=1$,
$H_2=2$, $s_r=4$, and $s_m=5$); 
the smaller grains are white and
the larger grains red. 
We find stratification and also reproduce the ``kink'' mechanism 
explained in the text
when we improve upon the simplest ``mean
field'' model by including four different angles of repose to take into
account the fact that the angle of repose depends on the concentration
of grains at the surface of the pile \protect\cite{makse}.}
\label{dynamics}
\end{figure}

\end{document}